# Reversible Metal-Semiconductor Transition of ssDNA-Decorated Single-Walled Carbon Nanotubes

Misun Cha,[1] Seungwon Jung,[2] Moon-Hyun Cha,[3] Gunn Kim,[4] Jisoon Ihm,[3] and Junghoon Lee[1, 5*]

[1]Institute of Advanced Machinery and Design, Seoul National University, Seoul 151-744, Korea

[2]School of Mechanical and Aerospace Engineering, Seoul National University, Seoul 151-744, Korea

[3]Department of Physics and Astronomy, Seoul National University, Seoul 151-747, Republic of Korea

[4]BK21 Physics Research Division and Department of Physics, Sungkyunkwan University, Suwon 440-746, Korea

[5]Interdisciplinary Program for Bioengineering, Seoul National University, Seoul 151-742, Korea

AUTHOR EMAIL ADDRESS * jleenano@snu.ac.kr

CORRESPONDING AUTHOR E-mail: jleenano@snu.ac.kr

ABSTRACT A field effect transistor (FET) measurement of a SWNT shows a transition from a metallic one to a p-type semiconductor after helical wrapping of DNA. Water is found to be critical to activate this metal-semiconductor transition in the SWNT-ssDNA hybrid. Raman spectroscopy confirms the same change in electrical behavior. According to our ab initio calculations, a band gap can open up in a metallic SWNT with wrapped ssDNA in the presence of water molecules due to charge transfer.

.

Recently, ssDNA-SWNT hybrids were introduced as intriguing bio-nano materials. Although they were initially used for dispersion in water,[1] new research directions are unfolded regarding the mechanism and changes in their physical properties resulting from the hybrid formation.[2-6] The ssDNA forms a stable complex with an individual SWNT through the aromatic interactions between nucleotide bases and SWNT walls, resulting in a wrapping configuration.[1] Force field calculations of SWNTs wrapped by ssDNA suggest that these hybrids are stable due to the π-staking between nucleic bases and the side wall of the SWNT.[1]

The change in the physical properties is a highly interesting phenomenon that results from the hybrid formation of the ssDNA-SWNT. This change has been observed with optical investigations such as Raman spectroscopy and UV/vis/NIR.[7-9] Near field photoluminescence showed a local spectra shift due to the wrapping of DNA.[10] On the other hand, direct measurements of the conductivity shift are rare, and those that have been achieved are inconclusive.[11] Furthermore, a detailed investigation of the changes in the transport properties due to the hybrid formation remains unexplored.

The present study represents the first direct measurement of the metal-semiconductor transition of a SWNT through a hybrid formation with ssDNA by helical wrapping. FET-type measurements were used to monitor the electrical properties after a dielectrophoretic deposition of the hybrid onto patterned electrodes. We found that water molecules were key elements for the activation of the transition. A reversible transition between the metallic and semiconducting behavior was demonstrated through repeated hydration and dehydration. It is well known that the electrostatic and conformational properties of DNA are sensitive to the surrounding water.[12] The polarity of the water molecules plays a crucial role in the neutralization and stabilization of DNA through charge screening of the negatively charged phosphate group. It can be expected that the electrical properties of ssDNA-SWNT hybrids are also influenced by the interaction between DNA and surrounding $H_2O$ molecules.

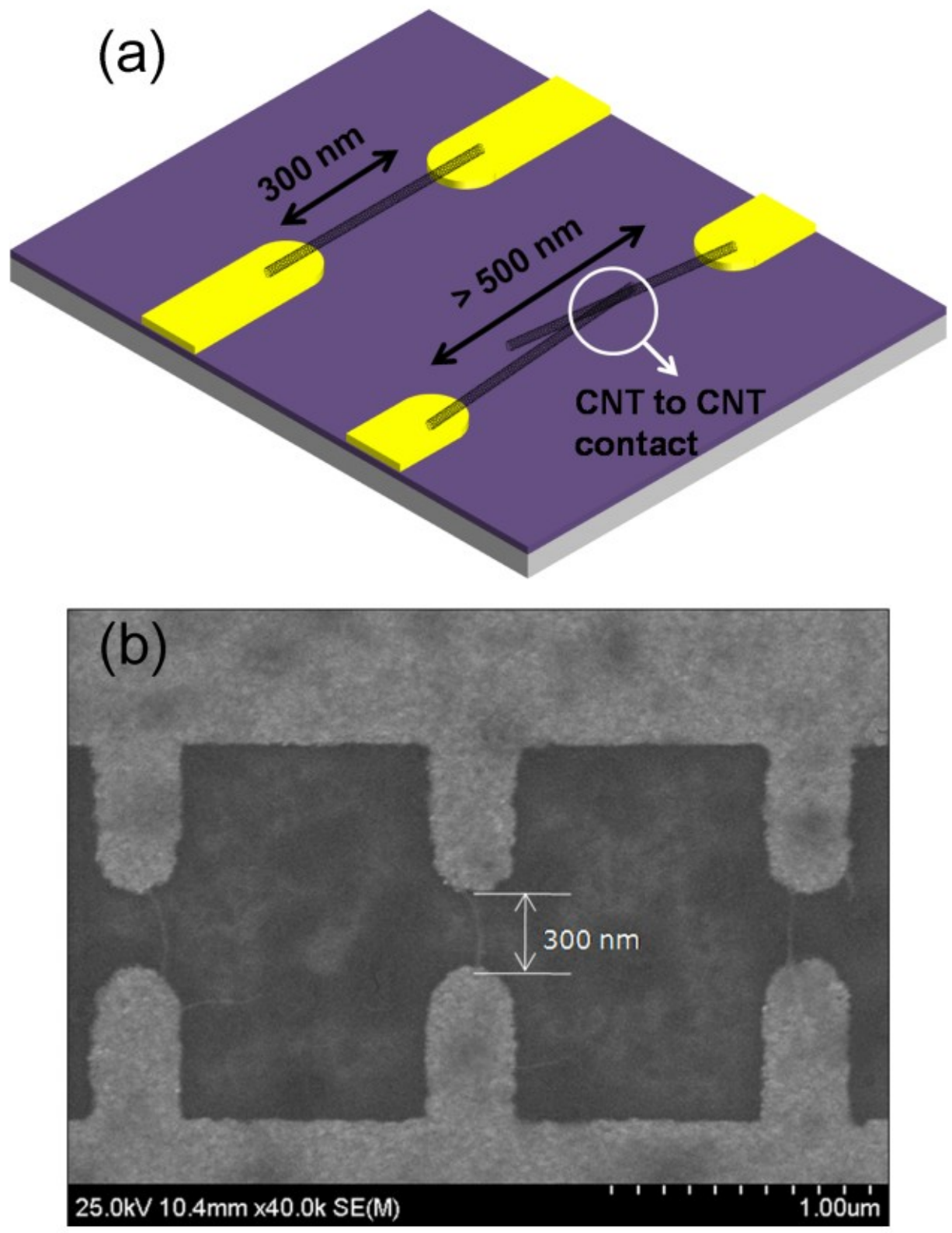

**Figure 1.**

We used a nanoscale gap device to measure the conductivity change due to the wrapping of DNA on SWNTs. The electrodes were fabricated by nanoimprint lithography (NIL) to facilitate the realization of a 300-nm gap. The CNT-to-CNT contacts were minimized since the gap size (300 nm) was smaller than the average length of the SWNTs (500 nm) as illustrated in Fig. 1a. The hybrids of ssDNA-SWNT were prepared using an established method.[13] Briefly, metallic dominant SWNTs (Carbon Nanotechnologies Inc. TX USA, Raman spectrum shown in Figure 1 in Supporting Information 1) synthesized through the high-pressure carbon monoxide conversion (HiPco) process were sonicated in an aqueous solution with 10 μM oligo-DNA (18-mer poly(C), and poly(G)) for 90 min to form ssDNA-SWNT hybrids.[12] The resultant suspension was centrifuged at 15,000 rpm for 60 min to remove large impurities in the sample. The sonication and centrifugation process were repeated twice. These processes ensure that no SWNTs remain without the formation of hybrids. The surfactant-based suspension of bare SWNTs used as a control sample was prepared in a similar process of sonication and centrifugation in a 1% sodium dodecyl sulfate (SDS) surfactant.

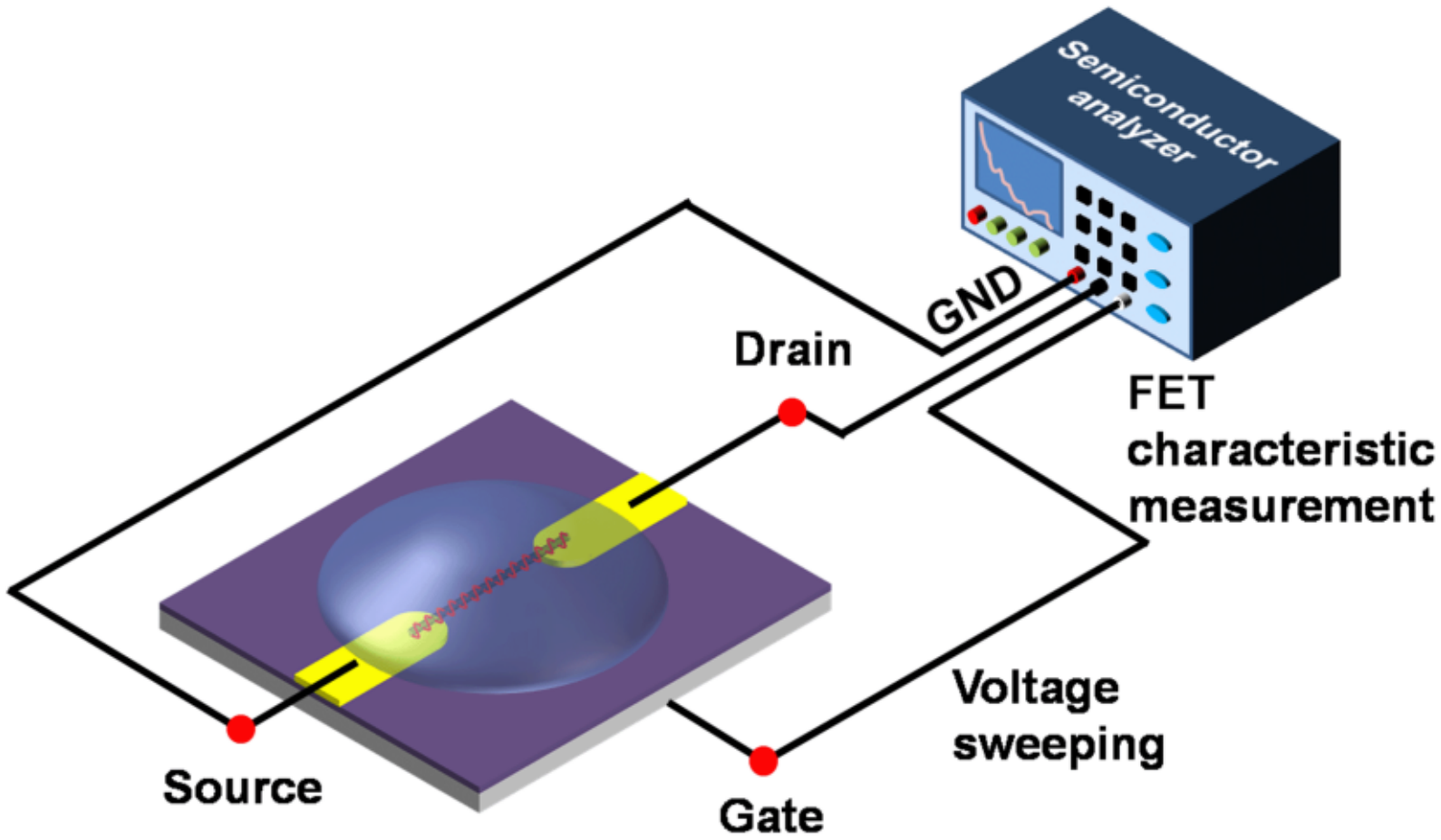

**Figure 2.**

The ssDNA-SWNT hybrids and the surfactant-based SWNTs were deposited by dielectrophoresis across pairs of electrodes.[14, 15] A 5 μl droplet of the solution prepared according to the protocol above was placed on the electrode. A voltage of 3 V at 5 MHz was maintained for 1 min, and then the sample was washed with DI water several times. This rinsing step ensures that the sample is free from salt and SDS which may affect the electronic structure of the SWNT. The sample was dried on a hot plate (~ 50 °C) for 5 min after the rinsing, which was followed by an inspection or experiments. [Geng, H.Z.; Lee, D.S.; Kim, K.K.; Han, G.H.; Park, H.K.; Lee, Y.H. Chem. Phys. Lett, 2008, 455, 275-278.] Figure 1b shows a scanning electron microscopy (SEM) image of ssDNA-SWNTs hybrids deposited on the electrodes. The deposition protocol was designed based on trial and error for the assembly of a countable number of SWNTs. For example, the SEM image shows that there are 1-2 bundles deposited across each gap connected in parallel. As there are 30 pairs of gaps, the total number of deposited bundles is less than 50. Dielectrophoretic deposition with the metallic dominant sample adds to the possibility of having more metallic SWNTs deposited [Krupke, R.; Henrich, F.; Löhneysen, H. v.; Kappes, M. M. Science, 2003, 301, 344-347]. Furthermore even when the SWNTs were wrapped with DNAs our DEP deposition promotes the deposition of metallic ones [Sickert, D.; Taeger, S.; Neumann, A.; Jost, O.; Eckstein, G.; Mertig, M.; Pompe, W. AIP Conf. Proc., 2005, 786, 271-274].

The electrical response of the ssDNA-SWNTs was measured as illustrated in Figure 2. The source-drain current ($I_{SD}$) was measured using a semiconductor analyzer (HP4155A) with the gate voltage swept from -15 V to +15 V.

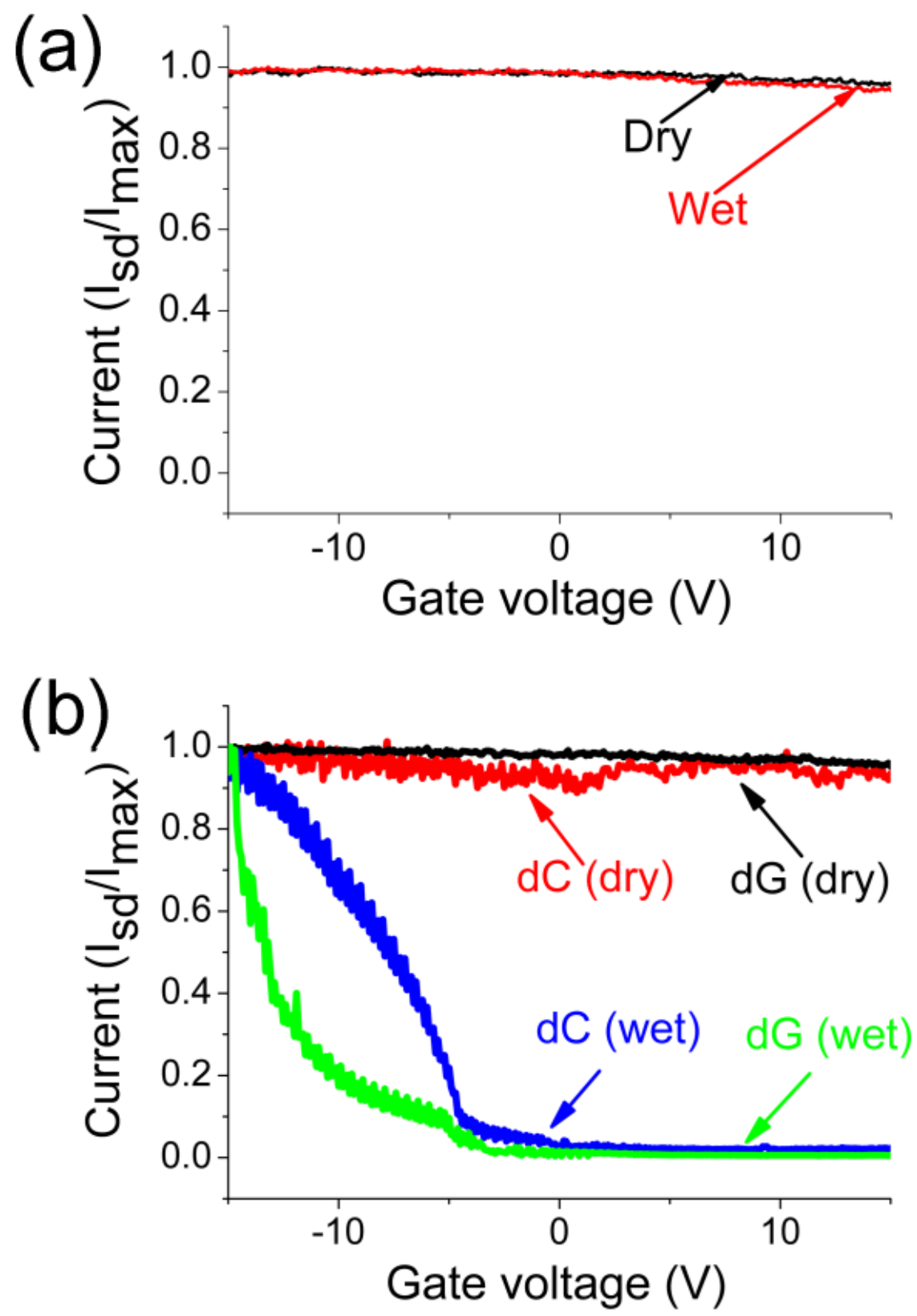

**Figure 3.**

Figure 3 shows an FET response normalized by the maximum value of the current. According to our observation the electrical property of the metallic SWNTs without DNA was not significantly affected by the presence or absence of water (Figure 3 a). [16] [Sung, D. ; Hong, Suklyun. ; Kim, Y.H.; Park N.; Kim S. ; Maeng S.L. ; Kim K.C. Appl. Phys. Lett, 2006, 89, 243110-243113] The same measurement with the ssDNA-SWNT hybrids in Figure 3b shows a sharp contrast. The source-drain current ($I_{SD}$) remained mostly flat in the dry state when the gate voltage was swept from -15 V to 15 V. The same sample in the wet state, however, manifests the behavior of a p-type semiconductor with the threshold voltage occurring at approximately -7 V. The current response characteristics are similar to previously reported FET measurements with semiconducting SWNTs.[18] Although there were some variations in detailed characteristics such as the threshold voltage and the maximum current, all measurements (> 20 times) with different samples always showed the p-type behavior. The detailed characteristics may change with variations in the deposition characteristics, but the distinctive behavior of the p-type semiconductor is always facilitated by the formation of the hybrid. This result demonstrates that the transport characteristic of the SWNT is significantly altered by the existence of DNA and water. Furthermore the current responses are close to zero above the cut-off or in the range of the positive gate voltage when the DNA wrapped SWNTs were tested (Figure 3 in Supporting Information 3). This observation reveals that the characteristic of metallic SWNT decreased significantly as a result of hydration. Thus we can exclude the possibility of property shift due to semiconducting SWNTs that may remain in the bundle without the metal-semiconductor transition.

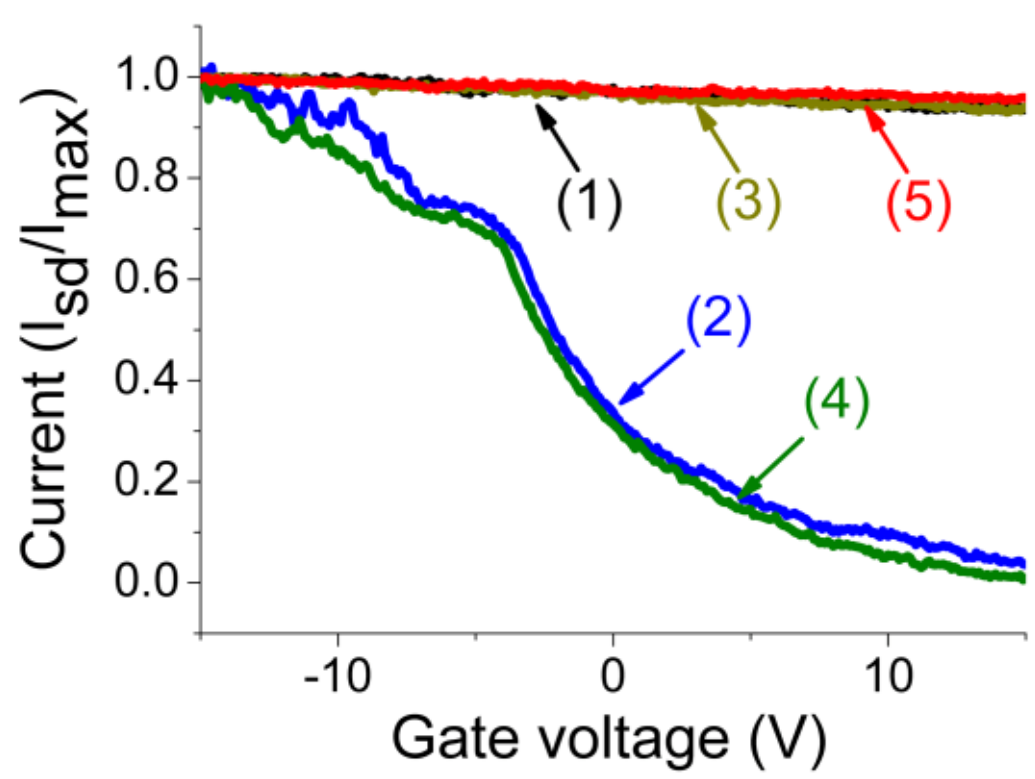

**Figure 4.**

To explore the influence of the DNA sequences, we also investigated the electrical characteristics of SWNT hybrids with heteropolymeric ssDNA molecules. As a heteropolymeric DNA, ssDNA: 5'-aaaggacgacattagacgaa-3' was selected. This oligonucleotide sequence is the specific region of *Staphylococcus aureus* 23s rDNA. In Figure 4, the FET characteristic of this hybrid shows a result similar to previous case involving a homopolymeric sequence; transition to a p-type semiconductor occurred due to the wrapping of the DNA with the heteropolymeric sequence and the addition of water. Figure 4 also shows that this metal-semiconductor transition is reversible when hydration and dehydration are repeated.

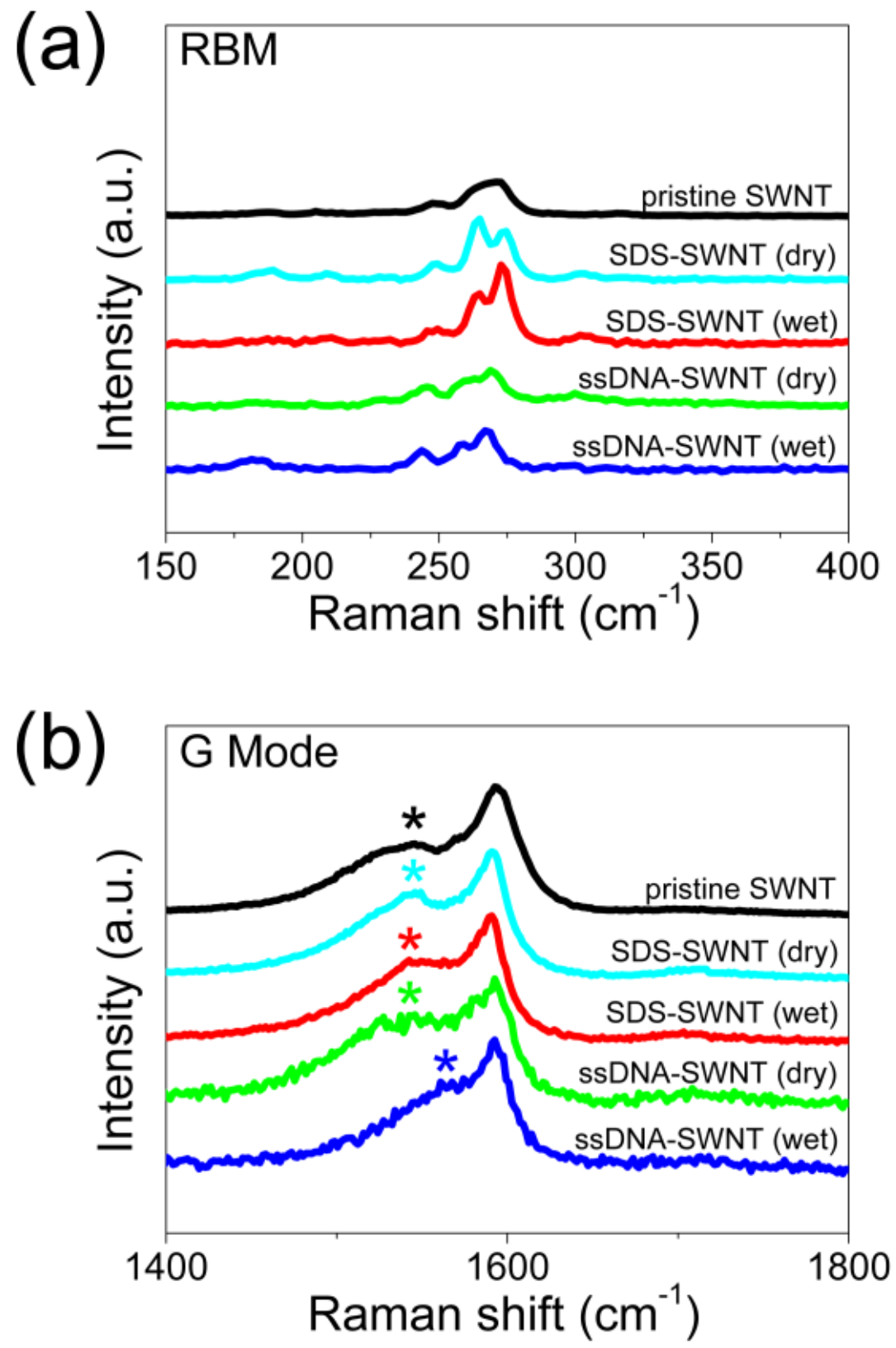

**Figure 5.**

Raman spectroscopy was used to further investigate the effect of water on the metal-semiconductor transition. Radial breathing modes (RBM) and tangential modes (G-bands) were measured in the Raman spectra of pristine SWNT in powder and ssDNA-SWNT hybrids on the electrodes. A micro-Raman system (Jobin-Yvon, LabRam HR) was used with laser lines at a wavelength of 514.5 nm from an Ar-ion laser for excitation. Figure 5a shows the RBM peaks that are used to determine the diameters of the samples. According to our analysis of the figure, the diameter appears to be $1.1 \pm 0.2$ nm for all three cases. Additionally the G-band shows a clear change in the electronic structure of the ssDNA-SWNT hybrid in water. In general, while the G-band in the Raman spectrum of semiconducting SWNT shows a sharp and symmetric Lorentzian line, that of metallic SWNT is represented by a broad and asymmetric Breit-Wigner-Fano (BWF) line shape ($G^-$ peak) as well as a Lorentzian line ($G^+$ peak).[19] Figure 5b shows the different G-bands of three types of SWNTs. The G-band of a pristine SWNT shows a broad and asymmetric BWF line shape. This is consistent with the result of the FET measurement shown in Figure 3a. The G-band of the deposited ssDNA-SWNT hybrid shows a similar BWF line in the dry state. After an injection of water into an identical sample, however, it was found that the optical properties changed considerably. The broad and asymmetric BWF line was greatly diminished and shifted to a high frequency. This kind of G-band shift due to charge carrier doing has been previously reported [Tsang]

First-principles electronic structure calculations of the complex of a metallic (6,6) SWNT and an adeno monophosphate (AMP) molecule were performed to further understand the experimental observations. Two model systems were chosen to mimic this experiment. While DNA is neutral since negatively charged phosphate groups are passivated by protons in a dry state, it is negatively charged in a wet state. For our computational models, a neutral AMP with hydrogen termination and a negatively charged AMP with a phosphate group surrounded by 11 $H_2O$ molecules were considered as shown in Figure 6. In the presence of water molecules, the phosphate group moves closer to the SWNT surface (with respect to the position of the P atom by 2.0 Å as shown in Figure 4 in Supporting Information 4). According to our Mulliken population analysis, there would be no practical charge transfer for a neutral AMP. In contrast, the (6,6) SWNT donates a fraction (~0.2 e) of an electron to negatively charged AMP surrounded by water molecules and becomes a p-type semiconductor (see Figure 4 in Supporting Information 4), which is in agreement with our experiment.

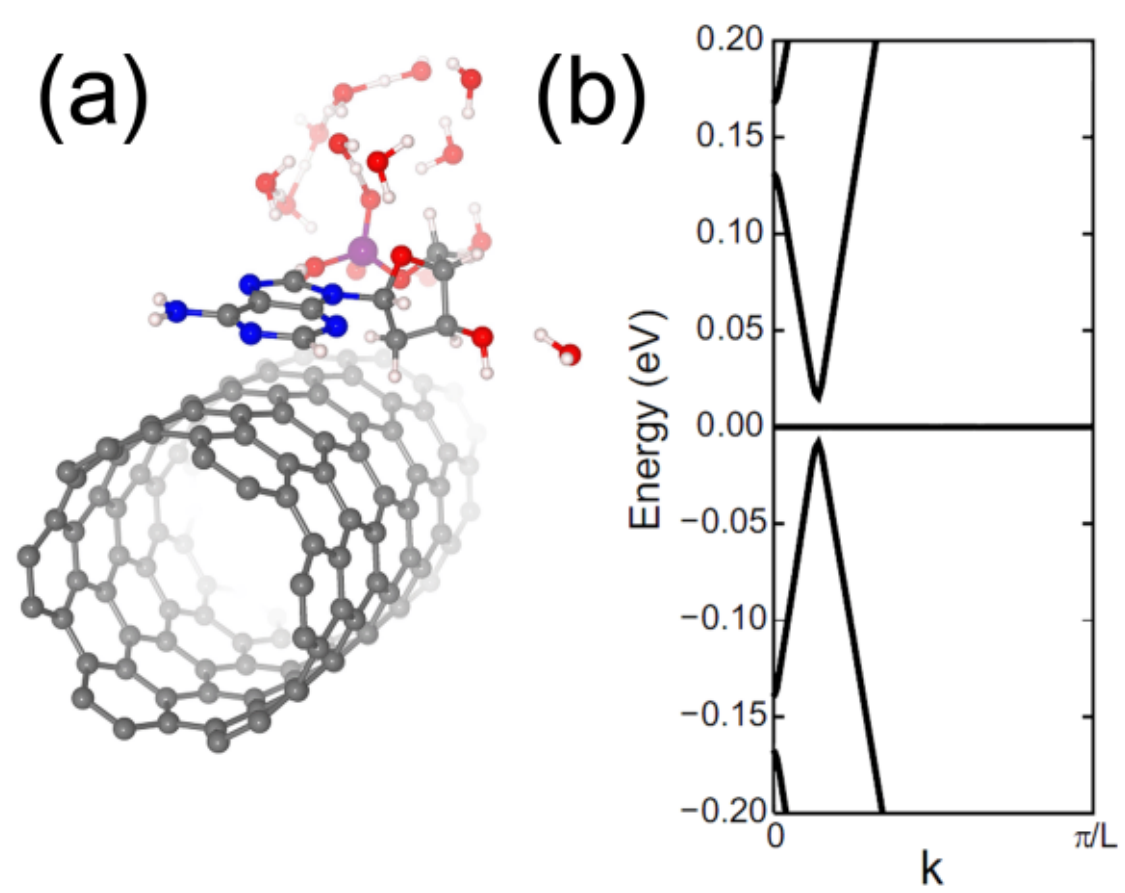

**Figure 6.**

According to our first-principles calculations, the energy band gap opens up by ~30 meV in the metallic (6,6) SWNT and the SWNT becomes a p-type semiconductor, as shown in Fig. 6. The electron transfer from the (6,6) SWNT to AMP and the subsequent formation of the positive image charges on the metallic SWNT surface due to the negatively-charged phosphate group give rise to local electric field and break the cylindrical symmetry, thus opens up a band gap. Consequently, the metallic armchair SWNT becomes a p-type semiconductor, which is in agreement with the present experiment. If ssDNA is introduced instead of a monomer (AMP in our model), the chiral and cylindrical symmetries of the metallic SWNT is broken by the helical perturbing potential caused by the ssDNA, which can be modeled as a charged helical line.[20] Even when the electron transfer between a ssDNA and a SWNT is not considered, a simple model of the helical perturbation can induce a band gap (a few meV to a few tens of meV)[20] in a metallic armchair SWNT. A synergistic effect from the electron transfer, the Coulomb interaction (between image charges and negative charged phosphate groups), and a helical perturbation of wrapped ssDNA can result in a more substantial opening of the gap. In the absence of water molecules, according to our computational results, negatively charged AMP would donate a fraction (~0.2 e) of an electron to the (6,6) SWNT (n-type doping), which is far from experimental results where DNA is neutralized in a dry state as mentioned above.

On the basis of experimental data and computational simulations, we conclude that the metal-semiconductor transition of ssDNA-decorated metallic SWNTs is caused by the charge transfer between the SWNT and ssDNA in water as well as by a helical perturbation resulting from ssDNA wrapping. It is expected that the properties of ssDNA-SWNT hybrids are influenced by the interaction between DNA and the surrounding water molecules. In the case of the ssDNA-SWNT hybrid, the bases bind to the surface of the SWNT through the π-stacking, exposing the hydrophilic sugar-phosphate backbone at the exterior.[1] When the ssDNA is hydrated with water molecules, hydration occurs more strongly around the phosphate backbone groups of the DNA, and the negatively charged phosphate group moves closer to the SWNT surface. Consequently, the electrons would transfer from the metallic SWNT to the negatively charged phosphate backbone, resulting in the transition to a p-type semiconductor with a band gap opening. These unique electronic properties with reversible metal-semiconductor transition of ssDNA-SWNT hybrids will provide critical information for applications based on DNA-SWNT hybrids such as nano-biosensing and gene/drug delivery.

## Methods

### Nano-size Electrode Fabrication by NIL

The mold was fabricated by electron-beam lithography (Lion-LV1, Leica) with reactive ion etching (RIE, P-5000, Applied Materials). A hydrophobic self-assembled monolayer (SAM, Heptadecafluoro-1, 1, 2, 2-Tetrahydrodecyl Trichlorosilane, Gelest) coating was used to ease the mold-releasing process. For the pattern transfer, aluminum was deposited by a thermal evaporator (MHS-1800, Muhan). Following this, Poly methyl metacrylate (PMMA, MicroChem) was coated onto a Si substrate by spin-coating. After coating the PMMA layer, MR I-8020 thermoplastic polymer (MicroChem) was spin-coated and baked. Imprinting was carried out at 190℃ and 45 bar for 20 minutes (Nanosys 420, Nano and Device). Reactive ion etching (RIE, RIE 80 Plus, Oxford) with oxygen gas was conducted at a power of 100 W and under a pressure of 55 mTorr. Aluminum wet etching was carried out using an aluminum etchant to obtain a negative slope. 5 nm Cr and 20 nm Au were deposited by a thermal evaporator at a rate of 0.4 Å/s (MHS-1800, Muhan). The resulting shape of the electrodes shows that electrodes with a gap of 300 nm were fabricated successfully (explanations in greater detail regarding the fabrication are given in Supplementary Information).

### Computational methods

Ab initio calculations were performed based on the density functional theory.[21] Wave functions were expanded in the double-zeta basis set implemented in the SIESTA package.[22, 23] Norm-conserving Troullier–Martins pseudopotentials[24] were employed. For the exchange–correlation term, the Ceperley–Alder type local density approximation[25] was employed with spin polarization, and the energy cutoff for real space mesh points was 250 Ry. All coordinates were fully relaxed until the Hellmann-Feynman forces of each atom were lower than 40 meV·Å$^{-1}$. The lateral (xy) dimension of the tetragonal supercell (with the nanotube axis along the z direction) is 35 Å, which is large enough to avoid intertube interactions. To calculate the electronic distributions in the systems, a Mulliken population analysis was done.

ACKNOWLEDGMENT: This work was supported by the Korea Research Foundation Grant funded by the Korean Government (MOEHRD) (KRF-J03000) and the Micro Thermal Systems Research Center under the auspices of the Korea Science and Engineering Foundation. Fabrication and experiments were performed at the Inter-university Semiconductor Research Center. We thank Dr. W. I. Choi for fruitful discussions and M. K. Choi of Raman Research Center. This work was also supported by the post BK21 project (M.-H.C. and G.K.) and the KOSEF grant (J.I.) funded by MEST (Center for Nanotubes and Nanostructured Composites). The computations were performed using the supercomputing resources of the Korean Institute of Science and Technology Information (KISTI).

SUPPORTING INFORMATION PARAGRAPH

Raman spectrum of the SWNTs used this study, nano-size electrode fabrication process using NIL and Raw data of FET characteristic of ssDNA-SWNT hybrids are available free of charge via the World Wide Web at http://pubs.acs.org.

FIGURE CAPTIONS

**Figure 1**. ssDNA-SWNT hybrids deposited on 300nm gap size electrode with DEP (a) Schematic of critical gap size of the electrode (b) SEM image of DEP deposition

**Figure 2**. Schematic of an experimental setup of a FET type measurement

**Figure 3**. Dependence of current of ssDNA-SWNT hybrid on gate voltage (a) A metallic behavior with SWNTs without DNA (b) A p-type semiconductor behavior with ssDNA-SWNT hybrid

**Figure 4**. Reproducibility tests; (1) initial dried sample, (2) first hydration, (3) first dehydration, (4) second hydration, (5) second dehydration (used sequence: 5’-aaa gga cga cat tag acg aa-3’)

**Figure 5**. Raman spectra of pristine SWNT and deposited ssDNA-SWNT hybrids in dry and wet state (a) RBM (b) G-band; The BWF line (black arrow) associated metallic SWNTs diminished after the injection of water.

**Figure 6**. AMP-SWNT hybrid with $H_2O$ molecules near a $PO_4^-$ group (a) Ball-and-stick model (b) Band structure of our model structure of AMP-SWNT hybrid with $H_2O$ molecules near a phosphate group.